\documentclass{ws-ijgmmp}

\begin{document}

\markboth{Iarley P. Lobo}
{On the physical interpretation of non-metricity in Brans-Dicke gravity}

%
\catchline{}{}{}{}{}
%

\title{On the physical interpretation of non-metricity in Brans-Dicke gravity}

\author{Iarley P. Lobo}

\address{Departamento de F\'isica, Universidade Federal da Para\'iba, C. Postal 5008, Jo\~ao Pessoa, PB 58051-970, Brazil\,\\
\email{iarley\_lobo@fisica.ufpb.br}}

\maketitle

\begin{history}
\received{(Day Month Year)}
\revised{(Day Month Year)}
\end{history}

\begin{abstract}
Brans-Dicke theory is described by an action that allows the so called frame transformation, which replaces the non-minimal coupling between the scalar field and the curvature by a coupling between the scalar field and matter fields. In this paper, we describe how the viewpoint that they are physically equivalent has a geometrical counterpart in the framework of Weyl integrable geometry. This way, Dicke's interpretation in terms of running units is a physical manifestation of the non-metricity tensor.
\end{abstract}

\keywords{Brans-Dicke theory; Frame transformations; Weyl geometry.}

\section{Introduction}
The fundamental symmetries of General Relativity (GR) have a geometrical meaning, for example the coordinate transformation invariance is described in a tensorial way, i.e., by an intrinsic definition of equations; the Lorentz invariance can be described by rotations and boosts of tetrad fields that keep invariant the local Minkowskian metric; and the theory is defined in a differential manifold endowed with a Riemannian metric that represents the gravitational field.
\par
The Brans-Dicke (BD) theory \cite{Brans:1961sx} is a scalar-tensor formalism that allows the intensity of the gravitational interaction to vary with the cosmic time and space location by turning Newton's constant into a field, which becomes also responsible for representing the gravitational field, along with the metric tensor. It is a scalar field, that not only is responsible for the coupling between matter and gravity, but also has its own evolution.
\par
An interesting property of the BD theory is the possibility of performing transformations in the metric and scalar fields that allow to recover the constant gravitational coupling and insert this information into a non-trivial form of the matter Lagrangian. It is still a controversial topic whether this transformation relates physically equivalent frames or not. In some cases they are treated as being physically inequivalent (for instance, see \cite{Bhadra:2006rn,Nozari:2012cy,Nozari:2009ds} and references therein), while in other cases it is adopted a viewpoint that permits to interpret them equivalently using a picture of running units \cite{Dicke:1961gz,Faraoni:2006fx,faraoni2012cosmology}. From this viewpoint, this transformation of {\it frames} is a symmetry of the theory, that does not exist in GR (which is just invariant under coordinate transformations). Despite of also being a symmetry, the frame transformation cannot have a geometrical interpretation in the language of Riemannian geometry.
\par
In this paper we exhibit such characterization in terms of a Weyl integrable geometry \cite{weyl1,Romero:2012hs}, which besides presenting the usual symmetries of GR, also presents a frame transformation symmetry and is able to describe the adaptations necessary to interpret them as equivalent using the geometrical tool of non-metricity. In the next section, we present the transformation that relates the Jordan to the Einstein frame and we present the interpretation of running units that permits to treat these as equivalent frames. In section \ref{w-geom} we review the Weyl geometry, and in section \ref{geom-frames} we demonstrate how Weyl geometry is a natural mathematical language for the interpretation that Jordan and Einstein frames are physically equivalent. In \ref{conc} we draw our concluding remarks.


\section{Change of Frames}\label{section_change_of_frames}
In the context of scalar-tensor theories of gravitation, in which we consider two dynamical fields that are responsible for the gravitational interaction, we have the prototype of these theories represented by the Brans-Dicke action 
\begin{equation}
S=\int d^4x\frac{\sqrt{-\text{det}\, g}}{16\pi}(\Phi R-\frac {\omega }{\Phi}g^{\mu \nu}\Phi _{,\mu}\Phi _{,\nu}+16\pi L_m), \label{brans-dicke}
\end{equation}
where $\text{det}\, g$ is the determinant of the metric $g_{\mu\nu}$, $R$ is the Riemannian Ricci scalar, $\Phi _{,\mu}\doteq \partial_{\mu}\Phi$ and $L_m$ is matter Lagrangian.
\par
This theory presents the so called non-minimal coupling of the curvature scalar with $\Phi$ represented in the action by the term $\Phi R$. In fact, we verify that the Newton's constant $G$ was replaced by the scalar field, which conceptually attends the objective of Carl H. Brans and Robert H. Dicke \cite{Brans:1961sx}, in 1961, in turning $G$ into a field, $G\mapsto 1/\Phi$, i.e., in treating the intensity of the gravitational coupling as a variable quantity, which depends on the spacetime point, thus being able to vary in the history of the Universe. This way, one would have a formulation that is more `Machian' than General Relativity.
\par
It was immediately realized in 1962 \cite{Dicke:1961gz} that by means of simple transformations in the metric tensor and scalar field, it was possible to remove the coupling of $\Phi$ with the curvature and return to the canonical form of General Relativity, with the presence of a constant $G$. These transformations are
\begin{subequations}
\begin{eqnarray}
\bar{g}_{\mu \nu}=G\Phi g_{\mu \nu},\\
\bar{\phi}=\sqrt{\frac{2\omega +3}{16\pi G}}\ln \left(G\Phi\right),
\end{eqnarray}
\end{subequations}
where $\omega >-3/2$. The new action assumes the GR form
\begin{equation}
S=\int d^4x\sqrt{-\text{det}\, \bar{g}}\left[\frac{\bar{R}}{16\pi G}-\frac{1}{2}\bar{g}^{\mu \nu}\bar{\phi}_{,\mu}\bar{\phi}_{,\nu}+\exp\left(-8\sqrt{\frac{\pi G}{2\omega +3}}\bar{\phi}\right)L_m(\bar{g}/G\Phi)\right].\label{transformedaction}
\end{equation}
\par
Notice that the non-trivial coupling of $\Phi$ and $R$ is transferred to the matter Lagrangian $L_m$ and the standard coupling of GR was restored.
\par
The first case, with non-minimal coupling defines the so called \textit{Jordan frame} (JF); while the second one defines the \textit{Einstein frame} (EF), because in this case, the action and equations of motion assume the General Relativity shape, just adding a scalar matter content.
\par
At this point one might wonder: which frame, in fact, corresponds to the physical reality? Faraoni discusses this issue in some details in Refs.~\cite{Faraoni:2006fx,faraoni2012cosmology} and we see that the major part of the literature on the subject takes Einstein frame as the physical one, due to the positively definiteness of the kinetic energy in the weak field limit and the existence of a ground state, see also Ref.~\cite{1994PhRvD..50.5039M} and references therein. Despite this good property of the EF, since in JF the worldlines of test particles are those that extremize the arc-length, i.e., are geodesics of a Riemannian geometry, in transforming to the Einstein frame one does not have anymore Riemannian geodesics (of metric $\bar{g}$) as their worldlines, but instead one has accelerated curves, which implies in a violation of the weak equivalence principle in the Einstein frame \cite{Faraoni:1999hp}, since they deviate from geodesics of $\bar{g}$ due to a force proportional to $\bar{\phi}$ \cite{Faraoni:2006fx}.
\par
The point that Faraoni raises to discussion is the existence of a third viewpoint, the one defended by Dicke himself in Ref.~\cite{Dicke:1961gz}: the third way is that {\it both frames are equivalent}.
\par
Following Dicke's reasoning, physical results cannot depend on the choice of units of measurement used and not only for the case in which one increases or decreases the scale of rules and clocks by means of a global transformation, but this should be a feature valid even if the rescaling change is point-dependent, i.e., even if it is constructed with a local transformation. A simple way of writing this idea is by choosing a scalar, smooth and positively defined function, which is responsible for the map between metrics, like the one that we used in the transformation between frames. By performing a conformal transformation on the metric, we can see that the line element is also transformed. 
\par
In fact, from the transformation $g_{\mu \nu}\mapsto \bar{g}_{\mu \nu}=\Omega^{2}(x)g_{\mu \nu}=G\Phi(x) g_{\mu \nu}$,
\begin{equation}
ds^2\mapsto d\bar{s}^2= \bar{g}_{\mu \nu}dx^{\mu}dx^{\nu}=g_{\mu \nu}\Omega^2(x) dx^{\mu}dx^{\nu}=\Omega ^2(x)ds^2
\end{equation}
\par
This map induces a redefinition of space and time coordinates and can be interpreted as a point-dependent redefinition of units. In fact, if one defines the basis (keeping the coordinates fixed)
\begin{eqnarray}
e^{a}(x)=\Omega(x) \delta^a_{\mu}dx^{\mu},\label{t-basis}\\
E_a(x)=\Omega^{-1}(x)\delta^{\mu}_a\partial_{\mu},
\end{eqnarray}
we see that the conformal metric $\bar{g}$ transforms as
\begin{equation}\label{t-metric}
\bar{g}_{ab}(x)=\Omega^{-2}\delta^{\mu}_a\delta^{\nu}_b\, \bar{g}_{\mu\nu}(x)=\delta^{\mu}_a\delta^{\nu}_b\, g_{\mu\nu}(x),
\end{equation}
thus keeping the same functional dependence on the coordinates $\{x^{\mu}\}$ as the original one. This basis is clearly non-holonomic, since
\begin{equation}
[E_a,E_b]h(x)\neq 0,
\end{equation}
where $h(x)$ is a scalar function. This way, using Eqs. (\ref{t-basis}) and (\ref{t-metric}), the transformed line element is
\begin{equation}
d\bar{s}^2=\bar{g}_{ab}e^ae^b=\Omega^2(x)g_{\mu\nu}dx^{\mu}dx^{\nu}=\Omega^2(x) ds^2.
\end{equation}
Therefore, it is possible to transfer the conformal transformation on the metric tensor (defined in the coordinate basis $\{\partial_{\mu}\}$) to the non-holonomic basis $\{E_a\}$ and keep the metric components unaltered.
\par
Physically this means that if we consider that rods and clocks do not have fixed scales, but may vary depending on the spacetime point by means of the field $\Omega^2(x)=G\Phi(x)$, we see that, under this interpretation, Jordan and Einstein frames are equivalent in the Brans-Dicke theory.\footnote{Other quantities suffer with this rescaling, like the mass of particles and matter fields.  But, fundamental constants like the speed of light, Planck's constant, electronic charge are invariants.} Since an experiment always measures a ratio between a certain quantity and an arbitrary unit at a point, this scaling performed by means of a scalar function that transforms the physical quantity and the arbitrary unit preserves this ratio, thus keeping invariant the measured quantity (see Eq.(3.3) of Ref.\cite{Faraoni:2006fx}).
\par
Summarizing this viewpoint: in Jordan frame, spacetime units are constant, while Newton's ``constant'' may vary; in Einstein frame, Newton's constant is indeed constant again and its previous JF's point dependence is transferred to the units that describe this interaction: space, time and matter.\footnote{An interesting generalization may be performed in the context of disformal transformations (see Ref.~\cite{Bittencourt:2015ypa,Carvalho:2015omv,Lobo:2017bfh} and references therein), which generalize conformal ones and do not have an equivalent interpretative counterpart.} In the next sections we will see how this physical interpretation has a geometrical counterpart in the language of Weyl geometry.


\section{Weyl geometry}\label{w-geom}
Weyl geometry is a simple generalization of Riemannian geometry, in which it allowed that under parallel displacements the norm of vectors may vary. This is achieved by the, so called, W-compatibility condition \cite{weyl1,Romero:2012hs,Almeida:2013dba,Pucheu:2016act,Lobo:2015zaa,Novello:1992tb,Pucheu:2015qsa,Avalos:2017wtp,Barcelo:2017tes,Scholz:2017pfo}
\begin{equation}
\nabla_{\alpha}g_{\mu\nu}=\sigma_{\alpha}g_{\mu\nu},
\end{equation}
where $\sigma_{\alpha}$ is a one-form field called {\it Weyl field}. This generalization defines a legitimate geometry, for example we can define the components of a connection that obeys such compatibility condition as
\begin{equation}
\Gamma^{\alpha}_{\mu\nu}=\{_{\mu\nu}^{\alpha}\}-\frac{1}{2}g^{\alpha\beta}[g_{\mu\beta}\sigma_{\nu}+g_{\nu\beta}%
\sigma_{\mu}-g_{\mu\nu}\sigma_{\beta}],
\end{equation}
where $\{_{\mu\nu}^{\alpha}\}$ are the Christoffel symbols of $g$. This geometry has the interesting property of having a larger symmetry group than Riemannian geometry, in this case one can perform the transformations
\begin{subequations}\label{weyl_trans}
\begin{align}
\bar{g}_{\mu\nu}=Fg_{\mu\nu},\\
\bar{\sigma}_{\mu}=\sigma_{\mu}+\frac{F_{,\mu}}{F},
\end{align}
\end{subequations}
where $F=F(x)$ is a scalar function. This means that one can perform a conformal transformation followed by a gauge transformation and the compatibility condition and some geometrical quantities (like the connection and curvature) are preserved. The different configurations $(g,\sigma)$ are called {\it frames}.
\par
The special case in which $\sigma$ is an exact form, i.e., $\sigma_{\mu}=\varphi_{,\mu}$, for a scalar function $\varphi$, defines a {\it Weyl integrable manifold}. In this case, using Weyl transformations, it is always possible to nullify the non-metricity and absorb it into the metric. In fact, by choosing in (\ref{weyl_trans}) $F=e^{-\varphi}$, which implies that $(g,\varphi)\mapsto (e^{-\varphi}g,0)$. The former is called {\it Weyl frame} (WF) and the latter {\it Riemann frame} (RF). In the integral case, the general gauge transformation (\ref{weyl_trans}) becomes
\begin{subequations}\label{weyl_trans2}
\begin{align}
\bar{g}_{\mu\nu}=Fg_{\mu\nu},\\
\bar{\varphi}=\varphi+\ln(F).
\end{align}
\end{subequations}


\section{Geometrization of frames}\label{geom-frames}
In this section, we show how these different frames of Brans-Dicke theory can be described using an intrinsic geometrical language by means of an identification of these different configurations with natural gauges of a Weyl geometry, in which the scalar field gains a geometrical interpretation. With this viewpoint, {\it the coupling constant that measures the intensity of gravitational interaction, $G$, which in Brans-Dicke theory  is a field, acquires the interpretation of measuring the non-metricity in a non-Riemannian spacetime.}
\par
First of all, let us remember that, like General Relativity, Brans-Dicke theory in Jordan frame is defined in a Riemannian geometry. According to the results of the last section, in a Weyl integrable geometry, which is basically a scalar-tensor formalism, it is always possible to move to a frame in which the compatibility condition is that of a Riemannian geometry, the Riemann frame. Therefore, we are allowed to treat BD theory in the Jordan frame as defined in such Riemann frame, i.e., JF=RF. Considering this ansatz, let us see to what kind of theory we are led to if we perform a geometrical Weyl transformation that ``turns on'' the Weyl field.
\par
If in action (\ref{brans-dicke}), we imagine that we are in a Riemannian frame $(g,0)$, and perform the Weyl transformation (\ref{weyl_trans2}) for $F(x)=G\Phi(x)$ we have
\begin{subequations}
\begin{align}
\bar{g}_{\mu \nu}= G\Phi g_{\mu \nu},\\
\bar{\varphi}=\ln \left(G\Phi\right).
\end{align}
\end{subequations}
Doing this, we see that the action (\ref{brans-dicke}) assumes the form:
\begin{equation}\label{preaction}
S=\int d^4x\frac{\sqrt {-\text{det}\, \bar{g}}}{16\pi}\left[\frac{1}{G}\bar{g}^{\mu \nu}\bar{R}^W_{\mu \nu}-\frac{\omega}{G}\bar{g}^{\mu \nu}\bar{\varphi}_{,\mu}\bar{\varphi}_{,\nu}+16\pi e^{-2\bar{\varphi}}L_m(e^{-\bar{\varphi}}\bar{g})\right].
\end{equation}
\par
We know from Weyl geometry that the connection components are invariant under gauge transformations (\ref{weyl_trans}), as a consequence, so is the Riemann tensor. Therefore, it is straightforward to conclude that the Ricci tensor $R_{\mu \nu}$ in the Riemannian frame $(g,0)$ has the same value as its counterpart in Weyl frame $(\bar{g}=G\Phi g,\bar{\varphi}=\ln(G\Phi))$, i.e., we mean that the tensor $\bar{R}_{\mu\nu}^W\doteq R_{\mu \nu}((G\Phi)^{-1}\bar{g})$ which appears in (\ref{preaction}) is equal to the Ricci tensor in Weyl geometry $\bar{R}^{W}_{\mu \nu}(\bar{g}, \bar{\varphi})$. This way, Brans-Dicke's action, under a Weyl transformation, assumes the known \textit{WIST} form \cite{Novello:1992tb}
\begin{eqnarray}
S=\int d^4x\sqrt {-\text{det}\, \bar{g}}\left[\frac{1}{16\pi G}\bar{R}^{W}-\frac{\omega}{16\pi G}\bar{g}^{\mu \nu}\bar{\varphi}_{,\mu}\bar{\varphi}_{,\nu}+e^{-2\bar{\varphi}}L_m(e^{-\bar{\varphi}}\bar{g})\right].
\end{eqnarray}
\par
To compare these transformations with those done in section \ref{section_change_of_frames}, we must be cautious and realize that the curvature scalar which appears in (\ref{transformedaction}) is Riemannian, i.e., it is built just with the Christoffel symbols for metric $\bar{g}$, while the above action has contributions from the Weyl field, therefore, in order to effectively compare both actions we must absorb them in the kinetic term.
\par
Referring to the Riemannian Ricci tensor as $\bar{R}_{\mu \nu}^R$, we can do the decomposition presented in Ref.~\cite{Novello:1992tb}:
\begin{equation}
\bar{R}_{\mu \nu}^W=\bar{R}_{\mu \nu}^R+\bar{\varphi}_{,\mu ||\nu}+\frac{1}{2}\bar{\varphi}_{,\mu}\bar{\varphi}_{,\nu}+\frac{1}{2}\bar{g}_{\mu \nu}(\bar{\Box}\bar{\varphi}-\bar{\varphi}_{,\lambda}\bar{\varphi}^{,\lambda}),
\end{equation}
and by contracting with $\bar{g}^{\mu\nu}$ we have
\begin{equation}
\bar{R}^W=\bar{R}^R+3\bar{\Box}\bar{\varphi}-\frac{3}{2}\bar{\varphi}_{,\mu}\bar{\varphi}^{,\mu},
\end{equation}
where $\bar{\varphi}_{,\mu ||\nu}$ is the pure Riemannian covariant derivative (in metric $\bar{g}$, of course). Discarding terms of Riemannian divergence $\sqrt{-\bar{g}}\bar{\Box}\bar{\varphi}$, we find\footnote{In Ref.\cite{Bhattacharya:2017pqc} there is a discussion about the relevance of the boundary term and its relation to the Gibbons-Hawking-York boundary term.}
\begin{eqnarray}
S=\int d^4x\sqrt {-\text{det}\, \bar{g}}\left[\frac{1}{16\pi G}\bar{R}^R-\frac{1}{2}\left(\frac{2\omega+3}{16\pi G}\right)\bar{g}^{\mu \nu}\bar{\varphi}_{,\mu}\bar{\varphi}_{,\nu}+e^{-2\bar{\varphi}}L_m(e^{-\bar{\varphi}}\bar{g})\right].
\end{eqnarray}
If we redefine
\begin{equation}
\bar{\phi}= \sqrt{\frac{2\omega +3}{16\pi G}}\bar{\varphi} \ ; \ e^{-\bar{\varphi}}=1/G\Phi,
\end{equation}
we find an action identical to Brans-Dicke's one in Einstein frame (\ref{transformedaction})
\begin{equation}
S=\int d^4x\sqrt {-\text{det}\, \bar{g}}\left[\frac{1}{16\pi G}\bar{R}_R-\frac{1}{2}\bar{g}^{\mu \nu}\bar{\phi}_{,\mu}\bar{\phi}_{,\nu}+\exp\left(-8\sqrt{\frac{\pi G}{2\omega +3}}\bar{\phi}\right)L_m({\bar{g}/G\Phi})\right].
\end{equation}
\par
This way, we are induced to conclude that the so called frame transformations in Brans-Dicke theory may acquire a natural interpretation in the perspective of Weyl's integrable geometry. Besides that, we see clearly the geometrization of the Newton's constant. That is, the issue of transformation of frames in Brans-Dicke theory may be stated in the language of Weyl geometry, that naturally endows frame transformation and since this transformation is a symmetry of this geometry, the equivalence between these frames can be stated not only physically, but also geometrically, and is such that {\it running units in Einstein frame has a mathematical counterpart of being a manifestation of non-metricity}, i.e., Dicke's interpretation furnishes the physical meaning of the non-metricity tensor. And the equivalence principle, that seemed to be lost in the Einstein frame can be recovered since its world-lines are still geodesics, but of the Weylian connection.


\section{Conclusion}\label{conc}
According to Faraoni \cite{Faraoni:2006fx,faraoni2012cosmology}, with respect to these transformations, great part of literature does not consider variable units of space-time-matter as the correct interpretation of frame transformations in Brans-Dicke theory, which could accommodate the equivalence of these different representations. It is considered a mapping of a Riemannian theory in the Jordan frame in another with the same geometrical character of being Riemannian, the Einstein frame, which would have strong physical consequences, like the violation of the weak equivalence principle.
\par
So, we defined the Jordan frame in a geometric Riemann frame in Weyl geometry (to use its natural gauge transformations) decoupled the scalar field and the curvature and reached the Einstein frame as a geometrical Weyl frame. In other words, there is an equivalence of languages (JF,EF)=(RF,WF).
\par
Considering that these transformations are intrinsic to this geometry, i.e., that geometrically those frames are equivalent, we see that Dicke's interpretation \cite{Dicke:1961gz} is naturally mathematized, where the image of running units in Einstein frame is just a physical manifestation of non-metricity. For example, Riemannian geodesics are mapped into Weylian geodesics and the equivalence principle is preserved.
\par
Some few last words about this theme is related to the fact that in article Ref.~\cite{Novello:1992tb} the introduction of Weyl geometry is justified by arguing that if a perturbation of rods and clocks of the system of units in Minkowski spacetime is a mechanism responsible for the Universe formation, this perturbation would be carried out by Weyl's field. Therefore, one can see that this interpretation that relates non-metricity and running units already exists, despite that apparently, a connection to Dicke's interpretation was not done.
\par
All this analysis is done under a classical point of view, but as is pointed by Faraoni \cite{faraoni2012cosmology}, the equivalence of frames does not seem to be preserved in one quantizes this theory, which under our perspective, indicates that Weyl symmetry is broken at the quantum level \cite{Nojiri:2000ja,Kamenshchik:2014waa}. 
\par
As a complement, concerning the behavior of classical singularities under frame transformations, we refer to \cite{Bahamonde:2016wmz} for and interesting viewpoint, in which singularities are absent in one frame and are present in another one in the context of $f(R)$ gravity, which might indicate a breakdown of this symmetry at the classical level. On the other hand, we indicate \cite{Almeida:2013dba} for a Weyl-invariant definition of scalars that preserves the nature of the spacetime with respect to singularities.

\section*{Acknowledgments}
I thank Eduardo Bittencourt and the anonymous referee for reading a previous version of this manuscript and suggesting several valuable improvements. IPL is supported by Conselho Nacional de Desenvolvimento
Cient\'ifico e Tecnol\'ogico (CNPq-Brazil) by the grant No.
150384/2017-3 and thanks Coordena\c c\~ao de Aperfei\c coamento de Pessoal de N\'ivel Superior (CAPES) for financial support.

\end{document}